\documentclass[conference]{IEEEtran}

\renewcommand\IEEEkeywordsname{Index Terms}
\usepackage{graphicx}
\usepackage[position=b]{subcaption}
\usepackage{tikz,xparse}
\usetikzlibrary{dsp,chains}
\usetikzlibrary{matrix}
\usepackage{mathptmx}
\usepackage{verbatim}
\usepackage{calc}
\usepackage{ifthen}
\usepackage{xifthen}
\usepackage{cancel}
\usepackage{bm}
\usepackage{verbatim}
\usepackage{multirow}
\usepackage{cite}
\usepackage{relsize}

\usepackage[nolist]{acronym} 
\usepackage{pgfplots}
\usetikzlibrary{arrows,shapes,graphs,graphs.standard,quotes,arrows.meta}
\pgfplotsset{compat=newest}
\usetikzlibrary{matrix}

\usepackage[hyphens]{url}

\usepackage[bookmarks=false]{hyperref}
\usepackage{units}
\usepackage{amsmath, amsbsy, amssymb, latexsym }
\usepackage{comment}
\usepackage[utf8]{inputenc}
\usepackage{xcolor}
\usepackage{enumitem}
\usepackage{algorithm}
\usepackage{algpseudocode}

\usepackage{etoolbox} 

\usepackage{flushend}

\tikzset{>=latex}

\makeatletter
\tikzset{circle split part fill/.style  args={#1,#2}{%
		alias=tmp@name, 
		postaction={%
			insert path={
				\pgfextra{%
					\pgfpointdiff{\pgfpointanchor{\pgf@node@name}{center}}%
					{\pgfpointanchor{\pgf@node@name}{east}}%
					\pgfmathsetmacro\insiderad{\pgf@x}
					\fill[#1] (\pgf@node@name.base) ([xshift=-\pgflinewidth]\pgf@node@name.east) arc
					(0:180:\insiderad-\pgflinewidth)--cycle;
					\fill[#2] (\pgf@node@name.base) ([xshift=\pgflinewidth]\pgf@node@name.west)  arc
					(180:360:\insiderad-\pgflinewidth)--cycle;            
}}}}}  
\makeatother  

\definecolor{mittelblau}{RGB}{0, 126, 198}
\definecolor{violettblau}{cmyk}{0.9, 0.6, 0, 0}
\definecolor{rot}{RGB}{238, 28 35}
\definecolor{apfelgruen}{RGB}{140, 198, 62}
\definecolor{gelb}{RGB}{255, 229, 0}
\definecolor{orange}{RGB}{244, 111, 33}
\definecolor{pink}{RGB}{237, 0, 140}
\definecolor{lila}{RGB}{128, 10, 145}
\definecolor{hellgrau}{RGB}{224, 224, 224}
\definecolor{mittelgrau}{RGB}{128, 128, 128}
\definecolor{dunkelgrau}{RGB}{80,80,80}
\definecolor{anthrazit}{RGB}{19, 31, 31}
\definecolor{darkgreen}{RGB}{34,139,34}
\colorlet{Mycolor1}{green!10!orange!90!}
    
\tikzset{ vndsSplit/.style={ shape=circle split, circle split part fill={black,red}, draw, inner sep=0pt,minimum size=12.5pt, rotate=90},
        vnds/.style={ shape=circle, fill=black, draw, inner sep=0pt,minimum size=5pt},
        cnds/.style={ shape=rectangle, fill=white, draw, inner sep=0pt,minimum size=5pt}, 
        vndGs/.style={ shape=circle, fill=green, draw, inner sep=0pt,minimum size=5pt},
        vndGsg/.style={ shape=circle, draw=black, fill=green, draw, inner sep=0.05pt,minimum size=5pt},
        vndRs/.style={ shape=circle, fill=red, draw, inner sep=0pt,minimum size=5pt},     
        vndRsr/.style={ shape=circle, draw=red, fill=red, draw, inner sep=0pt,minimum size=5pt},     
        vndsc/.style={ shape=circle, fill=black, draw, inner sep=0pt,minimum size=10pt},
        cndsc/.style={ shape=rectangle, fill=white, draw, inner sep=0pt,minimum size=7.5pt}, 
        vndGsc/.style={ shape=circle, fill=green, draw=black, inner sep=0pt,minimum size=10pt},
        vndRsc/.style={ shape=circle, fill=red, draw, inner sep=0pt,minimum size=10pt}
}

\IEEEoverridecommandlockouts

\renewcommand{\vec}[1]{\mathbf{#1}}

\newcommand{\Am}{\vec{A}}

\makeatletter
\newcommand{\vast}{\bBigg@{2}}

\makeatother
\begin{document}
	\begin{NoHyper}
		\title{Deep Learning-based Polar Code Design}

		\author{\IEEEauthorblockN{Moustafa Ebada, Sebastian Cammerer, Ahmed Elkelesh and Stephan ten Brink} \thanks{This work has been supported by DFG, Germany, under grant BR 3205/5-1 and BR 3205/6-1.}
			\IEEEauthorblockA{
				Institute of Telecommunications, Pfaffenwaldring 47, University of  Stuttgart, 70569 Stuttgart, Germany 
				\\\{ebada,cammerer,elkelesh,tenbrink\}@inue.uni-stuttgart.de
			}
		}

		\makeatletter
		\patchcmd{\@maketitle}
		{\addvspace{0.5\baselineskip}\egroup}
		{\addvspace{-0.78\baselineskip}\egroup}
		{}
		{}
		\makeatother

		\maketitle
		
		\begin{acronym}
			\acro{ECC}{error-correcting code}
			\acro{HDD}{hard decision decoding}
			\acro{SDD}{soft decision decoding}
			\acro{ML}{maximum likelihood}
			\acro{GPU}{graphical processing unit}
			\acro{BP}{belief propagation}
			\acro{BPL}{belief propagation list}
			\acro{LDPC}{low-density parity-check}
			\acro{HDPC}{high density parity check}
			\acro{BER}{bit error-rate}
			\acro{SNR}{signal-to-noise-ratio}
			\acro{BPSK}{binary phase shift keying}
			\acro{AWGN}{additive white Gaussian noise}
			\acro{MSE}{mean squared error}
			\acro{LLR}{log-likelihood ratio}
			\acro{MAP}{maximum a posteriori}
			\acro{NE}{normalized error}
			\acro{BLER}{block error rate}
			\acro{PE}{processing element}
			\acro{SCL}{successive cancellation list}
			\acro{SC}{successive cancellation}
			\acro{BI-DMC}{Binary Input Discrete Memoryless Channel}
			\acro{CRC}{cyclic redundancy check}
			\acro{CA-SCL}{CRC-aided successive cancellation list}
			\acro{BEC}{Binary Erasure Channel}
			\acro{BSC}{Binary Symmetric Channel}
			\acro{BCH}{Bose-Chaudhuri-Hocquenghem}
			\acro{RM}{Reed--Muller}
			\acro{RS}{Reed-Solomon}
			\acro{SISO}{soft-in/soft-out}
			\acro{PSCL}{partitioned successive cancellation list}
			\acro{3GPP}{$3^{\text{rd}}$ generation partnership project}
			\acro{eMBB}{enhanced Mobile Broadband}
			\acro{PCC}{parity-check concatenated}
			\acro{CA-polar codes}{CRC-aided polar codes}
			\acro{CN}{check nodes}
			\acro{PC}{parity-check}
			\acro{GenAlg}{Genetic Algorithm}
			\acro{AI}{Artificial Intelligence}
			\acro{MC}{Monte Carlo}
			\acro{CSI}{Channel State Information}
			\acro{PSCL}{partitioned successive cancellation list}
			\acro{5G-NR}{5G new radio}
			\acro{5G}{$5^{th}$ generation mobile communication}
			\acro{mMTC}{massive machine-type communications}
			\acro{URLLC}{ultra-reliable low-latency communications}
			\acro{SGD}{stochastic gradient descent}
			\acro{NN}{neural network}
			\acro{MI}{mutual information}		
			\acro{SPA}{sum product algorithm}
			\acro{DL}{Deep learning}
			\acro{RL}{reinforcement learning}
			\acro{MIMO}{multiple-input multiple-output}
\end{acronym}

\begin{abstract}
In this work, we introduce a deep learning-based polar code construction algorithm. 
The core idea is to represent the information/frozen bit indices of a polar code as a binary vector which can be interpreted as trainable weights of a neural network (NN).
For this, we demonstrate how this binary vector can be relaxed to a soft-valued vector, facilitating the learning process through gradient descent and enabling an efficient code construction. 
We further show how different polar code design constraints (e.g., code rate) can be taken into account by means of \emph{careful} binary-to-soft and soft-to-binary conversions, along with rate-adjustment after each learning iteration. 
Besides its conceptual simplicity, this approach benefits from having the \emph{``decoder-in-the-loop''}, i.e., the nature of the decoder is inherently taken into consideration while learning (designing) the polar code. 
We show results for belief propagation (BP) decoding over both AWGN and Rayleigh fading channels with considerable performance gains over state-of-the-art construction schemes.
			
\end{abstract}

\vspace{0.2cm}


\acresetall
\vspace{-0.2cm}

\section{Introduction}

\ac{DL} methods have been recently applied in many domains of communications. Examples span from trainable channel decoders \cite{nachmani2016learning,yihan2019deepturbo,kim2018communication,jiang2019mind,gruber2017deep,lyu2018performance,jiang2018learn}, \ac{NN}-based detectors, e.g., for \ac{MIMO} \cite{samuel2017deep} or for molecular channels \cite{farsad2017detection} up to end-to-end learning of a complete communication system \cite{oshea2017introduction}. Unfortunately, in particular in the field of channel (de-)coding, the exponential training complexity \cite{gruber2017deep} typically prevents from practical applications of a \emph{naive} learning approach (i.e., fully learned without any given structure; see \cite{gruber2017deep}).
Therefore, the authors of \cite{nachmani2016learning,kim2018communication,cammerer2017scaling} proposed to exploit the given code structure and pre-initialize the \ac{NN} structure which reduces the training significantly.
Hereby, the most common approach is to interpret the unrolled graph of the \ac{BP} decoder as a trainable \ac{NN} \cite{nachmani2016learning} which turns out to work well for short length codes even with (relatively) dense decoding graphs.

However, besides the implementation of \ac{DL} during \emph{run-time} (online complexity), i.e., the replacement or refinement of \emph{classical} signal processing components by \acp{NN}, \ac{DL} can be also used during \emph{design-time} (offline complexity), e.g., to find \emph{good} codes by the help of \ac{DL} tools as in \cite{jiang2018learn}. This work focuses on the later one, i.e., we use \ac{DL} to \emph{design} the polar code but use the classical decoders without any further modification. 

Although completely different, the previously mentioned approaches have in common that they 
optimize real-valued weights. However, in communications and, particularly, channel coding, many essential problems boil down to the task of finding binary representations that are omnipresent in today's digital communication systems. Thus, its worth mentioning that the proposed techniques from this work can be also applied to other communications (and signal processing) tasks.

On the other hand, polar codes \cite{ArikanMain} have recently attracted a lot of research as they  were selected to be among the error-correcting codes used in the control channel of the \ac{eMBB} service \cite{polar5G2018} in the \ac{5G} standardization process. They are also a potential candidate for \ac{URLLC} which faces the two major challenges of ultra high reliability and ultra low \emph{latency}. 
They are the first class of error-correcting codes theoretically proven to be asymptotically capacity-achieving under \ac{SC} decoding. However, the problem of finding polar decoders that fulfill the practical constraints (e.g., target latency, error-rate, hardware-complexity, etc.) in the finite-length regime is not yet fully solved. Furthermore, the problem of polar code design that fits other types of decoders that are of completely different decoding characteristics than \ac{SC} is still open. Nonetheless, the problem description is relatively \emph{simple} and from a \ac{DL} perspective the fundamental problem of polar code design is how to \emph{learn} a binary vector optimizing the error-rate performance for a given practical decoder. 

Polar codes are typically decoded by the \ac{SCL} decoder \cite{talvardyList}, while using a high-rate \ac{CRC} code to select the correct codeword from the list.
The increased implementation 
complexity (e.g., hardware complexity, memory requirements, etc.) of the \ac{SCL} decoder raised the urge for more simplified variants, e.g., \cite{AdaptiveList,PSCL_Journal}. However, the persistent drawback is the absence of soft-in/soft-out information 
processing, due to their sequential hard-decoding nature, which prevents their effective deployment in iterative detection/decoding schemes.
Besides \ac{SCL} decoding, polar codes can also be decoded with \emph{iterative decoders} \cite{ArikanBP}, which do not have these aforementioned problems. This class of decoders could be potentially suitable for high data rate applications (i.e.,  decoding throughput and \emph{latency}, from an implementation perspective). In \cite{elkelesh2018belief}, a \ac{BPL} decoder was introduced, with an error-rate performance close to \ac{ML}.
Unfortunately, existing variants of \ac{BP}-based decoders have an error-rate performance which is inferior to that of the CRC-aided \ac{SCL} decoder. However, due to their unique advantages, further performance enhancements (i.e., code properties and decoding performance) are worth considering. Existing polar code construction methods that are tailored to the \ac{BP} decoder appear to be few (cf. \cite{GenAlg_Journal_IEEE} and the discussion therein).


In this work, we propose a \emph{deep learning}-based polar code\footnote{Strictly speaking this is not an original polar code as no channel polarization is involved (cf. \ac{SC} decoding). However, it follows the same concepts and can be encoded/decoded alike.} construction algorithm. Following the same spirit as \cite{GenAlg_Journal_IEEE, GenAlgLDPCDecInTheLoop}, we \emph{learn} the polar code (i.e., frozen and non-frozen indices) for a given decoder and a specific channel. This showed to be a rather simpler code construction method  than explicitly finding the analytical solution for general channels while taking the decoder properties into account. 
Therefore, we follow the general idea of Nachmani et al. \cite{nachmani2016learning}, i.e., we apply iterative loop unrolling (\emph{deep unrolling} \cite{hershey2014deep}) and interpret the graph as a trainable \ac{NN}. However, unlike \cite{nachmani2016learning} we do not add any trainable weights to the graph itself but interpret the frozen/non-frozen positions as trainable weights of the \ac{NN}. As the unrolled graph is completely differentiable, we can apply \emph{standard} \ac{SGD}-based training of these weights with the only exception that the frozen positions itself need to be carefully transformed from a binary vector into a \emph{soft} one.

Opposed to \cite{jiang2018learn}, our approach explicitly makes use of a given code structure (i.e., the \emph{Hadamard} structure) and only \emph{learns} the location of the frozen bit-channels. The benefits are a reduced training complexity and the availability of well-understood \emph{classical} decoders. Therefore, our approach only introduces offline training complexity, however, once the design is fixed, the decoding complexity itself is the same as for any other \emph{classical} design scheme. The decoding complexity may even be reduced as less iterations, for a fixed error-rate performance, can be sufficient, especially if the code design is well-suited for the \ac{BP} algorithm. Also the channel and its noise statistics (e.g., \ac{AWGN} or Rayleigh) is inherently taken into account, as the \ac{SGD}-based training directly operates based on the actual data, rather than information theoretic approximations that may be inaccurate (e.g., the  Bhattacharyya bound for the \ac{AWGN} channel). Furthermore, this allows optimizing the code with respect to a fixed number of iterations and for any given finite block length.

\section{Polar Codes and BP decoding} \label{sec:con}		

Polar codes are conventionally based on the concept of channel polarization assuming that an \ac{SC} decoder is used.
As the code length $N=2^n$ gets larger, the resulting bit-channels become purely noiseless or purely noisy (i.e., \emph{polarize}). Afterwards, information and frozen (known) bits are transmitted over the noiseless and the noisy bit-channels, respectively.
The outcome from the polar code construction (or design) phase is the set of bit-channels used for information transmission. We denote the information and the frozen bit-channels by the sets $\mathbb{A}$ and $\bar{\mathbb{A}}$, respectively. Since a bit-channel is either used for information or frozen bit transmission, thus, merging the two sets (i.e., $\mathbb{A}$ and $\bar{\mathbb{A}}$) into one binary vector $\mathbf{A} = \left[ a_1,\dots,a_i,\dots,a_N \right] \in \mathbb{F}_2^N$ is straightforward. The bit-channel $i$ is described by the value of $a_i$ (e.g., $a_i = 1$ means that bit-channel $i$ is used for information transmission). 

Polar codes are based on a (polarization) matrix $\mathbf{F}$, an $N \times N$ matrix constructed based on the $2 \times 2$ Arıkan kernel $\mathbf{F}_2$ as
$ \mathbf{F} = \mathbf{F}_2^{\otimes n} $,
where $\mathbf{F}_2 = \left[ \begin{array}{ll} 1 & 0 \\ 1 & 1 \end{array}\right]$ and $\mathbf{F}_2^{\otimes n}$ denotes the $n$-th Kronecker power of $\mathbf{F}_2$.
Since polar codes are linear codes, encoding to find a codeword $\mathbf{x}$ is just a vector $\mathbf{u}$ by the matrix $\mathbf{F}$ multiplication. The vector $\mathbf{u}$ consists of the $k$ information bits in the $\mathbb{A}$ positions and $N-k$ frozen (known) bits. Throughout this work, we use the notation $\mathcal{P}(N,k)$-code to denote a polar code of length $N$ and code dimension $k$. 


\begin{figure}[t]
	\centering
	\resizebox{\columnwidth}{!}{	
		\begin{subfigure}{0.5\columnwidth}
			\resizebox{\columnwidth}{!}{\tikzset{text=black, font={\fontsize{14pt}{12}\selectfont}}

\let\pgfmathMod=\pgfmathmod\relax


\def\n{3}  \def\N{8}
\begin{tikzpicture}
\tikzset{edge/.style = {-, thick}}

\tikzset{h1/.style={preaction={
draw,yellow,-,
double=yellow,
double distance=4\pgflinewidth,
}}}

\tikzset{h2/.style={preaction={
draw,green,-,
double=green,
double distance=4\pgflinewidth,
}}}
          \pgfmathtruncatemacro{\nn}{(\n + 1)}
          
\foreach \i in {1,...,\N} {
\foreach \j in {1} {
	   \ifthenelse{\i=1 \OR \i=2 \OR \i=3 \OR \i=5}
	{\node [vndRsc] (v\i\j) at (2*\j ,9-\i) {};	}
	{\node [vndsc] (v\i\j) at (2*\j ,9-\i) {};	}
   }}

    \foreach \i in {1,...,\N} {
\foreach \j in {1,...,\n} {
   \ifthenelse{\j>1}
		{\node [vndsc] (v\i\j) at (2*\j ,9-\i) {};
       		\node [cndsc] (c\i\j) at (2*\j+1 ,9-\i) {};}      
       		{\node [cndsc] (c\i\j) at (2*\j +1,9-\i) {};} 		
   }}

\foreach \i in {1,...,\N} {
\foreach \j in {4} {
	\node [vndGsc] (v\i\j) at (2*\j ,9-\i) {};	
   }}

    \foreach \i in {1,...,\N} {
       \foreach \j in {1,...,\n} {
        		\draw (v\i\j)--(c\i\j);

   }}
   
      \foreach \i in {1,...,\n} {
                \pgfmathtruncatemacro{\ii}{(\i + 1)}
      \foreach \j in {1,...,\N} {
         		\draw (c\j\i)--(v\j\ii);

   }}

   \foreach \i [count=\xi] in {1,2,3} {
  \ifthenelse{\xi=1}{
  \draw (c1\i)--(v5\i);
   \draw (c2\i)--(v6\i);
   \draw (c3\i)--(v7\i);
   \draw (c4\i)--(v8\i);
   } {       \ifthenelse{\xi=2}{
     \draw (c1\i)--(v3\i);
   \draw (c2\i)--(v4\i);
   \draw (c5\i)--(v7\i);
   \draw (c6\i)--(v8\i);
   }{
     \draw (c1\i)--(v2\i);
   \draw (c3\i)--(v4\i);
   \draw (c5\i)--(v6\i);
   \draw (c7\i)--(v8\i);
     }
     }
}

\end{tikzpicture}
} \caption{Factor graph}
			\label{fig:convBP}
		\end{subfigure}   \hfill
	
	\begin{subfigure}{0.58\columnwidth}
		\resizebox{1\columnwidth}{!}{\tikzset{text=black, font={\fontsize{14pt}{12}\selectfont}}

\begin{tikzpicture}
\node[] (p1) at (0.4, -2.8) {}; 
\node[] (p2) at (1.6, -1.8) {}; 
\node[] (p3) at (0.4, 2) {}; 
\node[] (p4) at (1.6, 3) {};
\node[] (u1) at (-2, 2) {};
\node[] (u2) at (-2, -2) {};
\node[dspnodefull,minimum size=0.2mm] (temp) at (1, -2) {};
\node[dspsquare, minimum size=0.4cm](xor) at (1,2) {};\textsl{}
\node[dspnodefull, minimum size=0.4cm](xor2) at (1,-2) {};\textsl{}
\node[] (x1) at (4, 2) {};
\node[] (x2) at (4, -2) {};
\draw[line width = 0.5mm](u1)--node[above=0.1cm] {$R_{\mathrm{in,1}}$}(xor);
\draw[line width = 0.5mm](u1)--node[below=0.2cm] {$L_{\mathrm{out,1}}$}(xor);
\draw[dspconn,line width = 0.25mm]      (-1.2,2.25) --node[above] {} (0.3,2.25);
\draw[dspconn,line width = 0.25mm]      (0.3,1.75) --node[above] {} (-1.2,1.75); 
\draw[line width = 0.5mm](temp)--(xor);
\draw[line width = 0.5mm](xor)--node[above=0.1cm] {$R_{\mathrm{out,1}}$}(x1);
\draw[line width = 0.5mm](xor)--node[below=0.2cm] {$L_{\mathrm{in,1}}$}(x1);
\node[] (shiftxy) at (2.95, 3) {};
\draw[dspconn,line width = 0.25mm]      ($(shiftxy)+(-1.2,-0.75)$) --node[above] {} ($(shiftxy)+(0.3,-0.75)$);
\draw[dspconn,line width = 0.25mm]      ($(shiftxy)+(0.3,-1.25)$) --node[above] {} ($(shiftxy)+(-1.2,-1.25)$);
\draw[line width = 0.5mm](temp)--node[above=0.1cm] {$R_{\mathrm{out,2}}$}(x2);
\draw[line width = 0.5mm](temp)--node[below=0.2cm] {$L_{\mathrm{in,2}}$}(x2);
\node[] (shiftx) at (2.95, -1) {};
\draw[dspconn,line width = 0.25mm]      ($(shiftx)+(-1.2,-0.75)$) --node[above] {} ($(shiftx)+(0.3,-0.75)$);
\draw[dspconn,line width = 0.25mm]      ($(shiftx)+(0.3,-1.25)$) --node[above] {} ($(shiftx)+(-1.2,-1.25)$);
\draw[line width = 0.5mm](u2)--node[above=0.1cm] {$R_{\mathrm{in,2}}$}(temp);
\draw[line width = 0.5mm](u2)--node[below=0.2cm] {$L_{\mathrm{out,2}}$}(temp);
\draw[dspconn,line width = 0.25mm]      (-1.2,-1.75) --node[above=0.1cm] {} (0.3,-1.75);
\draw[dspconn,line width = 0.25mm]      (0.3,-2.25) --node[above=0.1cm] {} (-1.2,-2.25);

\draw[dspconn,line width = 0.25mm]      (0.75,-1.25) --node[above=0.1cm, rotate=90] {$L_{\mathrm{in,2}}+R_{\mathrm{in,2}}$} (0.75,1.25);
\draw[dspconn,line width = 0.25mm]      (1.25,1.25) --node[above=0.1cm, rotate=-90] {$f\left(R_{\mathrm{in,1}},L_{\mathrm{in,1}}\right) $} (1.25,-1.25);

\node[]  at (-1.85,2) {};
\node[]  at (-1.85,-2) {};
\node[]  at (3.85,2) {};
\node[]  at (3.85,-2) {};
\end{tikzpicture}} \caption{Processing element (PE)}
		\label{fig:PE}
	\end{subfigure}   \hspace{0.5cm}

	} 
    \caption{\footnotesize Original BP factor graph for a $\mathcal{P}(8,4)$-code with $\mathbb{A}=\{4,6,7,8\}$. Red nodes denote frozen positions and green nodes denote the LLRs of the channel output $\mathbf{L_{ch}}$.}
    \label{fig:GandHfactor}
	\vspace{-0.4cm}
\end{figure}
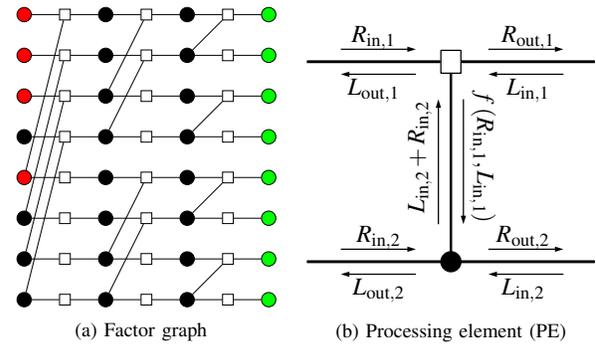

The \ac{BP} decoder of polar codes is a message passing decoder with an iterative decoding strategy. 
The underlying factor graph depends on the polarization matrix $\mathbf{F}$ of the polar code.
The frozen bits (i.e., the positions $\bar{\mathbb{A}}$ and the frozen values $\mathbf{u}_{\bar{\mathbb{A}}}$) are known to the decoder and this is used as the a priori information.
The channel output in terms of \ac{LLR} $\mathbf{L_{ch}}$ is the second input to the decoder. All other LLR messages are initialized to zero.

\ac{LLR}s are updated in the factor graph (Fig.~\ref{fig:convBP}) from left-to-right ($\bf{R}$-messages) and from right-to-left ($\bf{L}$-messages) according to equations (\ref{LReqn1})-(\ref{LReqn4}) till reaching a predefined maximum number of iterations $N_{it}$ or fulfilling a certain early stopping condition. A single  \ac{PE} is shown in Fig.~\ref{fig:PE}. The $\bf{L}$- and $\bf{R}$-messages are updated in each  \ac{PE} as follows:
\vspace*{-0.1cm}
\begin{eqnarray}\label{LReqn}
R_{\mathrm{out},1} &=f(R_{\mathrm{in},1},L_{\mathrm{in},2}+R_{\mathrm{in},2}) \label{LReqn1}
\\  
R_{\mathrm{out},2} &=f(R_{\mathrm{in},1},L_{\mathrm{in},1})+R_{\mathrm{in},2} \label{LReqn2}
\\    
L_{\mathrm{out},1} &=f(L_{\mathrm{in},1},L_{\mathrm{in},2}+R_{\mathrm{in},2}) \label{LReqn3}
\\  
L_{\mathrm{out},2}&=f(R_{\mathrm{in},1},L_{\mathrm{in},1})+L_{\mathrm{in},2}  \label{LReqn4}
\end{eqnarray}
where $f(x,y)=x \boxplus y$ is commonly referred to as \emph{boxplus} operator defined as 
\begin{equation*}
f(x,y)=x \boxplus y= \operatorname{ln}\frac{1+e^{x+y}}{e^x+e^y} .
\end{equation*}

After some user-defined number of iterations $N_{it}$, the \acp{LLR} that represent the estimated information vector $\hat{\mathbf{u}}$ and estimated codeword vector $\hat{\mathbf{x}}$ are given as: 
\vspace{-0.2cm}
\begin{eqnarray*}
L\left(\hat{u}_{i}\right)&=L_{i,1}+R_{i,1} 
\\ 
L\left(\hat{x}_{i}\right)&=L_{i,n+1}+R_{i,n+1}. \label{eqL}
\end{eqnarray*}
\begin{figure*}[t]
	\resizebox{\linewidth}{!}{\input{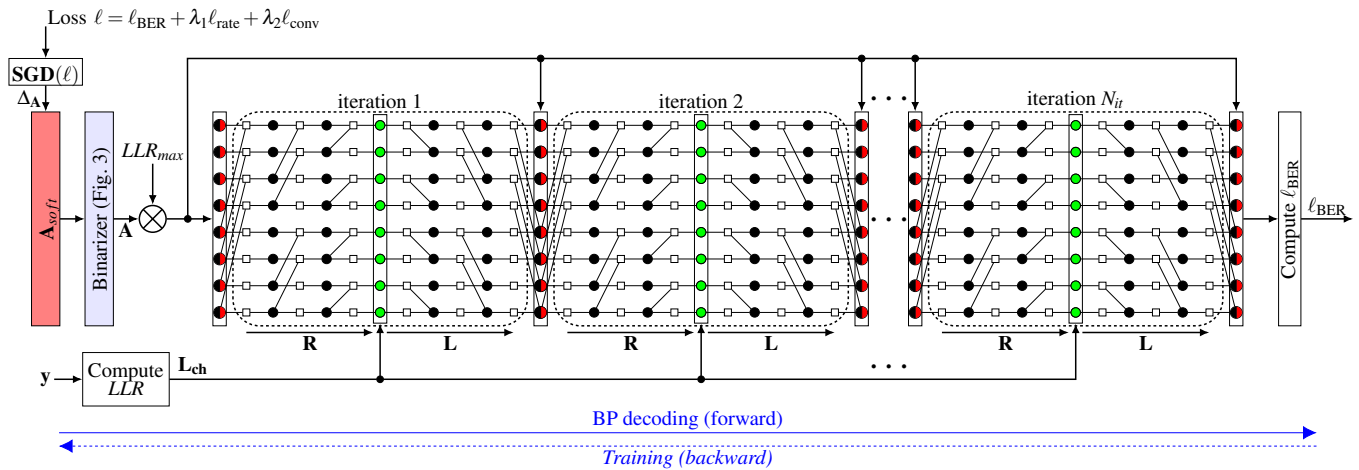}}  
	
	\vspace{-0.2cm}
	
	\caption{\footnotesize Learning framework over the unrolled BP factor graph of a polar code with codelength $N=8$ (i.e., a $\mathcal{P}(8,k)$-code).}
	\vspace{-0.55cm}	\label{fig:unrolledBP}		
\end{figure*}

\vspace{-0.6cm} Besides, \ac{BP} decoding can be performed over permuted factor graphs which greatly enhances the error-rate performance over both \ac{BEC} and \ac{AWGN} channel \cite{Urbanke_chCsC_BP,multi_trellis,elkelesh2018belief}. 
The polar code factor graph can be transformed into a \ac{LDPC}-like Tanner graph and, thus, \ac{LDPC} \ac{SPA} decoding is possible after applying some graph-pruning techniques \cite{sparseBP}.


Throughout this work, we use the \ac{BP} decoder over the encoding polar factor graph (Fig. \ref{fig:convBP}) with the flooding update rule. We use a fixed number of \ac{BP} iterations $N_{it}$ without any further early stopping condition.

\section{Code Construction Mismatch Paradigm} \label{sec:genAlg}

The state-of-the-art polar code constructions (e.g., \cite{ArikanMain,constructDE,constructGaussian,constructTalVardy,BetaIngmard,UPO}) assume that a hard-output \ac{SC} decoder is used. 
Thus, these algorithms are not optimal for iterative decoding and list decoding \cite{BP_LLR_Siegel,RMurbankePolar,polarDesign5G}.
The \ac{BP} decoder has a different decoding strategy when compared to the \ac{SC} decoder. A proper design tailored to \ac{BP} decoding should take into consideration the \ac{LLR} update schedule, the number of \ac{BP} iterations and the underlying specific permutation of the polar code factor graph. These parameters directly give an indication about the cycles, stopping and trapping sets in the graph.

In \cite{BP_LLR_Siegel}, the authors track the evolution of the \ac{LLR}s in the polar code factor graph during \ac{BP} decoding to find some bad information bit-channels and some good frozen bit-channels. Then these bit-channels are swapped enhancing the error-rate performance of the \ac{BP} decoder.
In \cite{GenAlg_Journal_IEEE}, a \ac{GenAlg}-based method is used which takes into account the real decoding behavior (i.e., “decoder-in-the-loop” concept). The resulting polar codes are tailored to both the considered decoder and the channel. It was observed that different polar codes are optimal depending on the considered decoder.
In \cite{ArikanBP}, it was shown that using the \ac{RM} rule leads to codes with poor performance under \ac{BP} decoding, thus, the minimum distance of the code $d_{min}$ is not the (only) parameter to be maximized in order to design codes suitable for iterative decoding. In this paper, we focus on designing polar codes tailored to iterative decoding using some deep learning techniques, but extensions to \ac{SCL}-tailored polar codes are possible.

Most of the available polar code construction techniques are assuming a \ac{BEC} or an \ac{AWGN} channel. However, designing a polar code, e.g., for the Rayleigh fading channel is an interesting on-going research problem in the coding community. Our proposed algorithm depends on real Monte-Carlo error-rate simulations at a specific channel condition and, thus, can be used to efficiently construct polar codes tailored to the Rayleigh channel at a certain design SNR.
	
\section{Learning to Construct polar codes}

As shown in Fig.~\ref{fig:unrolledBP}, we apply iterative loop unrolling \cite{nachmani2016learning,hershey2014deep} and replace the frozen/non-frozen bit vector $\Am$ by trainable weights. All \ac{BP} operations are differentiable and, thus, a \emph{standard} \ac{SGD}-based training can be applied to find the best weights with the only exception that the frozen positions vector itself should be a binary vector. To overcome this \emph{soft-to-binary} conversion, we re-interpret the real-valued weights as the probability that a specific position is frozen and evaluate it based on random realizations of $\Am$. With some modifications in the backward-propagation of \ac{SGD}, this method allows to \emph{learn} the frozen bit positions.
Further, our approach shares the same advantages as in \cite{nachmani2016learning}, and, as such, training with the all-zero codeword is possible. In the following, we assume without loss of generality that a frozen position is set to 0. 
For details on the loop unrolling for \ac{BP} decoding, we refer the reader to \cite{nachmani2016learning} and, in particular, for polar codes to \cite{doan2018neural}.

\subsection{From ``classical'' to stochastic code constructions}

The fundamental problem is that the optimization target (here the frozen/non-frozen positions $\Am$) is a binary decision and, therefore, a position is either frozen or non-frozen. In the following, we introduce a \emph{soft} trainable frozen/non-frozen position vector $\Am_\textrm{soft}$ that simply denotes the \ac{LLR}s (cf. \emph{logits} in the deep learning domain) of the probability that a position is frozen\footnote{Strictly speaking the \ac{BP} decoder would also work with soft-$\Am$ inputs, however, this does not lead to any meaningful interpretation at the encoder side as the code itself needs to be binary.}.

A common method for \emph{hard threshold} activations (e.g., a binary neuron) in deep learning literature is the so-called \emph{straight-through estimator} \cite{hinton2012coursera}. This simply means that the gradient during back-propagation is set as if the forward-direction had been the identity function. To further speed-up the convergence, we follow the approach from \cite{hubara2016binarized} where the authors propose a stochastic \emph{binarizer} layer, i.e., the samples of $\Am$ within a batch are randomly assigned with probability $p_i = \sigma(A_{soft,i})$ of being non-frozen. Hereby $\sigma(.)$ denotes the \emph{sigmoid} function.

The binarizer-layer behaves as an identity function in the backward-direction while in the forward-direction the output is
$$A_{i} = \begin{cases}
    +1     & \textrm{with probability } p_i = \sigma(A_{soft,i})\\
    0       & \textrm{with probability } 1-p_i.
\end{cases}$$
This procedure ensures that the resulting code construction $\Am$ describes one valid realization of a conventional binary polar code. Note that the same hard $\Am$-vector is used multiple times during decoding (after each iteration) and, thus, needs to be the same realization after each iteration as shown in Fig.~\ref{fig:unrolledBP}.
As a result, $p_i=\sigma(A_{soft,i})$ denotes the probability that bit-position $i$ is non-frozen and, thus, the average coderate is given by \vspace{-0.1cm}
\begin{equation}
R_{c,avg} = \frac{1}{N} \sum_{i=1}^N \sigma(A_{soft,i}).\label{eq:rate}
\vspace{-0.1cm}\end{equation}

The choice of the right loss function has to ensure that the training converges to the desired design rate. However, after training, the $N-k$ largest values of $\Am_\textrm{soft}$ are considered as frozen positions to ensure that the final code rate exactly matches the target rate and the code itself is deterministic.

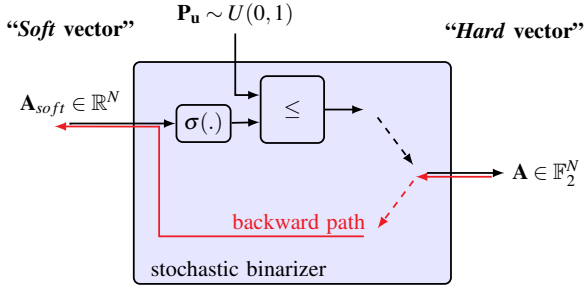
\begin{figure}[t]
	\centering
	\resizebox{0.9\columnwidth}{!}{\begin{tikzpicture}
\tikzset{>=latex}
\tikzstyle{box} = [fill= white!90!blue, draw,rounded corners=.1cm,minimum height=3em, minimum width=2em, align=center, thick] 
\tikzstyle{vertical_box} = [fill= white!90!blue,draw,rounded corners=.1cm,minimum height=2em, minimum width=3em, align=center, thick,rotate=90]

\node[box,minimum width=5cm,minimum height =3.5cm] (binarizer) at (0.,0) {};
\node[box,minimum width=1cm] (comp) at (0,1) {$\leq$};

\node[box,minimum width=0.5cm,minimum height = 0.5cm] (sigmoid) at (-1.4,0.78) {$\sigma(.)$};

\node[] (noise) at (-0.9,2.5) {$\mathbf{P_u} \sim U(0,1)$};
\node[] (Ain) at (-3.5,1.09) {$\Am_{soft} \in \mathbb{R}^N$};
\node[] (Aout) at (4,0) {$\Am \in \mathbb {F}_2^N$};

\node[] (l1) at (-0.85,-1.5) {stochastic binarizer};

\draw[->,thick] (noise) |- ([shift={(-0.04,0.4)}]comp);
\draw[->,thick] (Ain) |- (sigmoid);
\draw[->,thick] (sigmoid) -- ([shift={(0.0,-0.2)}]comp.west);

\node[] (out1) at (1.25,1) {};
\node[] (out2) at (1.25,-1) {};
\node[] (out3) at (2.0,0) {};
\node[] (int) at (-2.1,0.87) {};

\draw[thick,rot] (out2) -| (int);
\draw[->,thick,rot] ([shift={(0.0,-0.015)}]int.south) -- ([shift={(-0.25,-0.03)}]Ain.south);
\draw[->,thick,rot] ([shift={(-0.2,-0.06)}]Aout.west) -- ([shift={(-0.18,-0.06)}]out3.east);

\draw[->,thick] (comp.east) -- (out1);
\draw[->,thick,dashed] (out1) -- (out3);
\draw[->,thick,dashed,rot] (out3) -- (out2);

\draw[->,thick] (out3) -- (Aout);
\node[rot] (l2) at (0.1,-0.8) {backward path};

\node[] (hard) at (3.5,2.25) {\textbf{``\emph{Hard} vector''}};
\node[] (soft) at (-3.5,2.25) {\textbf{``\emph{Soft} vector''}};

\end{tikzpicture}}
	\caption{\footnotesize Block diagram of the proposed stochastic binarizer.}
	\label{fig:binarizer}		
	\vspace{-0.6cm}
\end{figure}

\subsection{Loss function and training methodology}

After having defined the basic setup, the detailed choice of the hyperparameters is given in Tab.~\ref{tab:hyperparameters} which essentially follows the common choice in the literature (cf. \cite{nachmani2016learning,gruber2017deep,jiang2018learn,oshea2017introduction}). Further gains may be achievable by extensive optimization of these hyperparameters, e.g., by means of heuristic search algorithms such as evolutionary algorithms \cite{jaderberg2017population}.
However, one crucial hyperparameter is the \emph{design} of the right loss function suited to the polar code design problem. We propose the weighted addition of multiple individual losses to fulfill all constraints, which are:
\begin{enumerate}
\item \emph{Low error-rates}. A good selection of the $\Am$-vector should minimize the error-rate after decoding either for information bits in $\mathbf{u}$ or codeword bits in $\mathbf{x}$. Thus, the binary cross-entropy loss is used 
\begin{equation*}
\begin{array}{llll}
\ell_\textrm{BER}(\mathbf{A}) &= &\frac{1}{k}\mathlarger{\mathlarger{\mathlarger{\sum}}}_{i \in \mathbb{A} }\vast( -(u_i)\log(\sigma(L(\hat{u}_i))) \\ 
 & &\,\,\,\,\,\,\,\,\,\,-(1-u_i)\log(1-\sigma(L(\hat{u}_i)))\vast).
\end{array}
\end{equation*}

 Alternatively, the \ac{MI} of the decoder output \acp{LLR} could be maximized (e.g., based on the \ac{MI} estimation in \cite{hoeher2000log}).

\item \emph{Target rate / rate adjustment}. It turns out to be beneficial if the code rate can be adjusted during training (i.e., first force convergence to higher rate codes and afterwards successively freeze more and more positions; similar to \emph{slack} variables). Let $R_\textrm{target}$ denote the desired code rate and with (\ref{eq:rate}) it follows
\vspace{-0.2cm}
\begin{equation*}
\begin{array}{ll}
\ell_\textrm{rate}(\Am_\textrm{soft},R_\textrm{target}) &= (R_{c,avg} -R_\textrm{target})^2 \\ 
&\stackrel{(\ref{eq:rate})}{=} \left( \frac{1}{N} \sum_{i=1}^N \sigma(A_{soft,i}) - R_\textrm{target} \right)^2.
\end{array}
\end{equation*} \vspace{-0.2cm}


\item \emph{Convergence towards deterministic behavior}. Although the training process is done based on stochastic realizations of $\Am$, the final output of the training algorithm should be a deterministic solution (i.e., \emph{the} learned code $\Am_{opt}$) and, thus, $A_{\textrm{soft},i} \to \pm \infty$. Therefore, we propose the additional loss
\vspace{-0.2cm}
$$\ell_\textrm{conv}(\Am_{\textrm{soft}}) = \frac{1}{N} \sum e^{-|A_{\textrm{soft},i}|}.$$

\item \emph{Avoiding local minima}. Unfortunately, the proposed losses are partly contradictive (e.g., the all-frozen vector obviously minimizes the \ac{BER} but does not fulfill the rate requirement). Thus, we introduce weighting factors $\lambda_i$ to adjust the individual impact per loss component during the training progress (e.g., in the beginning a relaxed rate loss penalty may be tolerable to allow more degrees of freedom).
\end{enumerate}
This leads to the final loss function: $$\ell = \ell_\textrm{BER}(\Am) + \lambda_1 \ell_\textrm{rate}(\Am_\textrm{soft},R_\textrm{target}) + \lambda_2 \ell_\textrm{conv}(\Am_\textrm{soft})$$
where $\lambda_i$ denotes the additional training hyperparameters.

Now, training is conducted in three phases:
\begin{itemize}
	\item Phase 1 (\emph{Initialization}), where an intermediate number of epochs is required to push the possibly randomly/weakly initialized $\mathbf{A}$-vector $\mathbf{A}_{\text{init}}$ towards a reasonably proper one. In this phase, $\lambda_1$ and $\lambda_2$ are set to 0, i.e., a relaxed loss penalty, as only BER acts as a loss in this phase and, thus, more degrees of freedom are provided.
	\item Phase 2 (\emph{Optimization}), where a reasonably large number of epochs is required to train the $\mathbf{A}$-vector \emph{given} the target code rate. In this phase, $\ell_\textrm{rate}$ is activated to constrain the training of the $\mathbf{A}$-vector by the target rate. Balancing both losses $\ell_\textrm{BER}(\Am)$ and $\ell_\textrm{rate}(\Am_\textrm{soft},R_\textrm{target})$ can be adjusted by careful choice of $\lambda_1$, while keeping $\lambda_2$ set to 0 (i.e., non-deterministic values of $\Am_{\textrm{soft}}$ are still allowed in this phase).
	\item Phase 3 (\emph{Saturation}), where a large number of epochs is required to \emph{fine-tune} the $\mathbf{A}$-vector by pushing the soft probabilities towards their extreme values. In this phase, $\ell_\textrm{conv}(\Am_\textrm{soft})$ is activated and allowed to dominate the loss function to push towards a more deterministic solution.
\end{itemize}

Phase 2 and 3 can be combined without a significant performance loss with the drawback of more difficult hyperparameter tuning.
For a given number of BP iterations $N_\textrm{it}$, training is performed over $2\cdot N_\textrm{it}\cdot (n+1) $ layers, where the factor of 2 accounts for the 2-way propagation in the iterative decoding algorithm.


\subsection{Approaches for non-differentiable decoders}

As \ac{SGD} relies on the availability of the gradient per layer, the proposed idea is limited to \ac{BP}-based decoders. However, the general idea of the soft $\Am_\textrm{soft}$-vector in combination with the straight-through estimator can also be applied with reinforcement learning (e.g., see \cite{aoudia2018end}) and, thus, also works for non-differentiable decoders.
This will facilitate learning on \ac{SCL}-based decoders for a specific list size over a user-defined channel state. The authors in \cite{GenAlg_Journal_IEEE} showed that this is critical, as a \emph{plain} \ac{SCL}-tailored polar code was able to achieve the same error-rate performance of a CRC-aided polar code under \ac{SCL} decoding. This relieves some complexity and brings back the pure structure of polar codes \emph{without} requiring a CRC/parity-check code in the loop.

		
\section{Results and Discussion}

In this work, training was performed for a $\mathcal{P}(256,128)$-code under BP decoding limited to $N_{\text{it}}=5$ iterations (i.e., corresponding to an equivalent of 90 NN-layers). This limitation was induced by some complexity constraints (e.g., huge size of the NN and its memory requirements) and, thus, pushed forward for a future study. However, gains were still observed for a slightly larger number of iterations.
The results for training over AWGN and Rayleigh fading channels are presented next. Extension of the current framework to accommodate different parameters (e.g., different codelengths, channel types, decoders, etc.)  is straightforward. In this work, we focus on the BP decoder-tailored design which, to the best of our knowledge, still misses a systematic/analytical approach.

\subsection{AWGN channel results}
We learn (i.e., \emph{design}) a $\mathcal{P}(256,128)$-code under BP decoding with BER as the error-rate cost function (i.e., $\ell_{\text{BER}}$). The newly obtained code is compared to the 5G polar code without the CRC-aid \cite{polar5G2018} under BP decoding with $N_{\text{it}}=5$ iterations. As depicted in Fig.~\ref{fig:BER_AWGN_comb}, the BER performance of the \emph{learned} code outperforms the performance of the 5G polar code under Arıkan's conventional BP decoder with $N_{\text{it}}=5$ iterations, leading to a performance gain in terms of BER. However, the gains are reduced for $N_{\text{it}}=20$ iterations due to the aforementioned training complexity constraints.

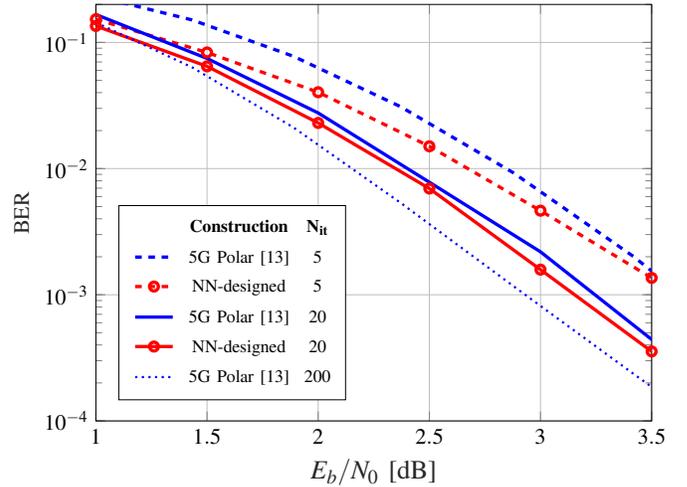
\begin{figure}[t]
	\centering
	
	\resizebox{\columnwidth}{!}{
%
%
\begin{tikzpicture}

\begin{axis}[%
width=\columnwidth,
height=0.75\columnwidth,
scale only axis,
xmin=1,
xmax=3.5,
xlabel style={font=\color{white!15!black}},
xlabel={\large $E_b/N_0$ [dB]},
ymode=log,
ymin=0.0001,
ymax=0.2,
yminorticks=true,
ylabel style={font=\color{white!15!black}},
ylabel={BER},
axis background/.style={fill=white},
title style={font=\bfseries},
xmajorgrids,
xminorgrids,
ymajorgrids,
mark options={solid},
legend style={at={(0.03,0.03)}, anchor=south west, legend cell align=left, align=left, draw=white!15!black}
]

\addplot [color=blue, dashed, line width=1.5pt]
table[row sep=crcr]{%
	1	0.232512934110371\\
	1.43299946165632	0.150693805183914\\
	1.88872205947323	0.077618565300286\\
	2.36968547635921	0.0310206238188445\\
	2.87885049445184	0.00924160052753596\\
	3.41973148833631	0.00200137753687863\\
	3.99654366295697	0.00030623828125\\
	4.61440275098179	3.63375e-05\\
	5.27960133120302	3.05670121192865e-06\\
	6	2.44629884769539e-07\\
};
	\label{lab:AWGN_5G_5Iter}


\addplot [color=red, dashed, line width=1.5pt,mark=o]
table[row sep=crcr]{%
	1	0.15266220703125\\
	1.5	0.08326103515625\\
	2	0.04028623046875\\
	2.5	0.01501513671875\\
	3	0.00463955078125\\
	3.5	0.0013603515625\\
};
	\label{lab:AWGN_NN_5Iter}

\addplot [color=blue, line width=1.5pt]
  table[row sep=crcr]{%
1	0.1663859375\\
1.5	0.07465546875\\
2	0.0276\\
2.5	0.0078\\
3	0.00219\\
3.5	0.0004421875\\
4	0.0001171875\\
4.5	7.8125e-06\\
5	9.375e-06\\
5.5	0\\
6	0\\
};
	\label{lab:AWGN_5G_20Iter}


\addplot [color=red, line width=1.5pt, mark=o, mark options={red}]
table[row sep=crcr]{%
	1	0.134978125\\
	1.5	0.06463046875\\
	2	0.023003125\\
	2.5	0.00695390625\\
	3	0.00158203125\\
	3.5	0.00035546875\\
};
	\label{lab:AWGN_NN_20Iter}

\addplot [color=blue, dotted, line width=1pt]
table[row sep=crcr]{%
	1	0.143811976578318\\
	1.43299946165632	0.0634307919552175\\
	1.88872205947323	0.0211869279661017\\
	2.36968547635921	0.00537011103343891\\
	2.87885049445184	0.00116830570792511\\
	3.41973148833631	0.000235157318669631\\
	3.99654366295697	4.3653120917902e-05\\
	4.61440275098179	6.74139613790579e-06\\
	5.27960133120302	8.46476187941712e-07\\
	6	7.19041499423866e-08\\
};
	\label{lab:AWGN_5G_200Iter}
%

\coordinate (legend) at (axis description cs:0.45,0.05);

\end{axis}

\matrix [
draw,
fill=white,
matrix of nodes,
anchor=south east,
font=\footnotesize,
mark options={solid}
] at (legend) {
	                          &  \bfseries{Construction}    & $\mathbf{N_{it}}$  \\
	\ref{lab:AWGN_5G_5Iter}   &  5G Polar \cite{polar5G2018}       & 5         \\
	\ref{lab:AWGN_NN_5Iter}   &  NN-designed    & 5         \\
	\ref{lab:AWGN_5G_20Iter}  &  5G Polar \cite{polar5G2018}       & 20        \\	
	\ref{lab:AWGN_NN_20Iter}  &  NN-designed    & 20        \\
	\ref{lab:AWGN_5G_200Iter} &  5G Polar \cite{polar5G2018}       & 200       \\
};

\end{tikzpicture}%
} 
	\vspace{-0.6cm}
	
	\caption{\footnotesize BER performance of the \emph{learned} 
		$\mathcal{P}(256,128)$-code over the \ac{AWGN} channel under BP decoding with number of iterations $N_{it}$.}\label{fig:BER_AWGN_comb}	
\end{figure}	
\begin{table}
	\vspace*{0.1cm}
	\caption{\footnotesize Hyperparameters used for training the $\mathcal{P}(256,128)$-code under BP decoding  limited to $N_{\text{it}}=5$ BP iterations }\label{tab:hyperparameters}
	\centering
	\begin{tabular}{{l} r}
		Parameter & Value \\
		\hline
		Optimizer &  \ac{SGD} with Adam \\
		Learning rate $L_r$ & 0.001 \\
		Training SNR $E_b/N_0$ & 2 dB, 4 dB and 5 dB \\
		Trainable parameters & 256 \\
		Effective number of layers & 90 \\
	\end{tabular}
	
\end{table}

\subsection{Rayleigh fading channel results}

There exist limited amount of work on Rayleigh-specific polar codes (e.g.,\cite{TrifonovRayleigh}). Next, same as before, a learned $\mathcal{P}(256,128)$-code is compared to the 5G polar code without the CRC-aid \cite{polar5G2018} under BP decoding with $N_{\text{it}}=\{5,20\}$ iterations. The system model is given by:
\vspace{-0.2cm}
$$y=\alpha \cdot x + n$$
where $y$ is the channel output, $n$ is the Gaussian noise attributed to the channel such that $n\sim\mathcal{N}(0,\sigma_{ch}^2)$ and $\alpha > 0$ is the fading coefficient which follows a Rayleigh distribution with $E[\alpha^2]=1$.
Assuming perfect \ac{CSI} (i.e., $\alpha$ is known to the  receiver at each received bit position), the corresponding channel \acp{LLR}, namely $L_{ch}$, are computed as 
\vspace{-0.2cm}
$$L_{ch}(y)=\log\frac{P\left(y|\alpha,x=+1\right)}{P\left(y|\alpha,x=-1\right)}=\frac{2}{\sigma_{ch}^2}\cdot \alpha \cdot y.$$

As depicted in  Fig.~\ref{fig:BER_ray_comb}, the BER performance of the \emph{learned} code outperforms the performance of the 5G polar code  under Arıkan's conventional BP decoder for both scenarios with $N_{\text{it}}=\{5,20\}$ iterations in terms of BER. 

As a reference, in both Fig.~\ref{fig:BER_AWGN_comb} and Fig.~\ref{fig:BER_ray_comb}, we compare the performance of the obtained codes to that of the 5G polar code under Arıkan's conventional BP decoder with $N_{\text{it}}=200$ iterations. It can be seen that the learned codes below $N_{it}=20$ iterations can possibly benefit from a reduced decoding latency and complexity (i.e., reduced number of BP iterations) without any loss in error-rate performance.

%
%
%
%
%
%
%

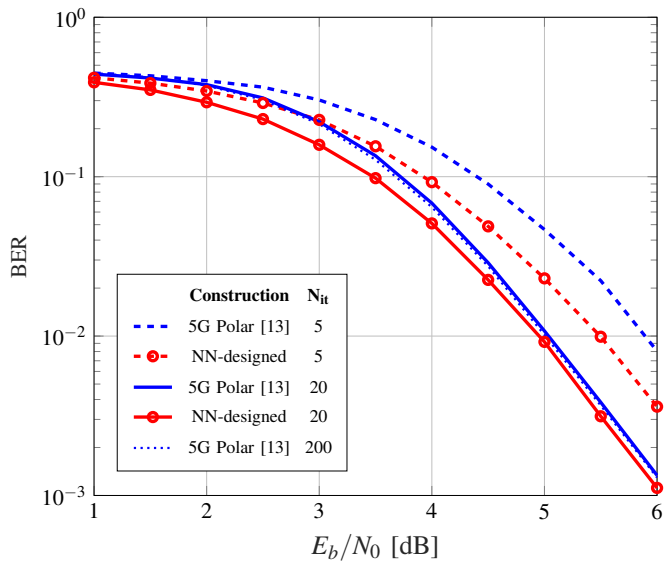
\begin{figure}[t]
	\centering
	
	\resizebox{\columnwidth}{!}{
%
%
\begin{tikzpicture}

\begin{axis}[%
width=\columnwidth,
height=0.85\columnwidth,
scale only axis,
xmin=1,
xmax=6,
xtick={1,2,3,4,5,6},
xlabel style={font=\color{white!15!black}},
xlabel={\large $E_b/N_0$ [dB]},
ymode=log,
ymax=1,
ymin=1e-3,
yminorticks=true,
ylabel style={font=\color{white!15!black}},
ylabel={BER},
axis background/.style={fill=white},
title style={font=\bfseries},
xmajorgrids,
xminorgrids,
ymajorgrids,
mark options={solid},
legend style={at={(0.03,0.03)}, anchor=south west, legend cell align=left, align=left, draw=white!15!black}
]

%

\addplot [color=blue, dashed, line width=1.5pt]
  table[row sep=crcr]{%
1	0.44654765625\\
1.5	0.4296\\
2	0.4000609375\\
2.5	0.36437421875\\
3	0.3028859375\\
3.5	0.22800546875\\
4	0.153203125\\
4.5	0.08986875\\
5	0.04661015625\\
5.5	0.0222609375\\
6	0.0080515625\\
};
\label{lab:Ray_5G_5Iter}


\addplot [color=red, dashed, line width=1.5pt, mark=o]
table[row sep=crcr]{%
	1	0.4173140625\\
	1.5	0.38643671875\\
	2	0.34462734375\\
	2.5	0.28997109375\\
	3	0.22717890625\\
	3.5	0.155115625\\
	4	0.09237578125\\
	4.5	0.0487921875\\
	5	0.023015625\\
	5.5	0.00990625\\
	6	0.00360859375\\
};
\label{lab:Ray_NN_5Iter}

\addplot [color=blue, line width=1.5pt]
table[row sep=crcr]{%
	1	0.44023671875\\
	1.5	0.41565234375\\
	2	0.378028125\\
	2.5	0.3119078125\\
	3	0.22094296875\\
	3.5	0.1352578125\\
	4	0.06824921875\\
	4.5	0.028834375\\
	5	0.0107671875\\
	5.5	0.0038421875\\
	6	0.0013375\\
};
\label{lab:Ray_5G_20Iter}


\addplot [color=red, line width=1.5pt, mark=o]
table[row sep=crcr]{%
1	0.390834375\\
1.5	0.35025\\
2	0.2929328125\\
2.5	0.22979296875\\
3	0.15846015625\\
3.5	0.09789453125\\
4	0.05091171875\\
4.5	0.02251640625\\
5	0.0091984375\\
5.5	0.00313984375\\
6	0.001115625\\
};
\label{lab:Ray_NN_20Iter}

\addplot [color=blue, dotted, line width=1pt]
table[row sep=crcr]{%
	1	0.4462546875\\
	1.5	0.4201765625\\
	2	0.3728296875\\
	2.5	0.3031625\\
	3	0.2138046875\\
	3.5	0.1275546875\\
	4	0.06441015625\\
	4.5	0.02740078125\\
	5	0.0101625\\
	5.5	0.0036671875\\
	6	0.001295\\
};
\label{lab:Ray_5G_200Iter}

\coordinate (legend) at (axis description cs:0.45,0.05);

\end{axis}

\matrix [
draw,
fill=white,
matrix of nodes,
anchor=south east,
font=\footnotesize,
mark options={solid}
] at (legend) {
	&  \bfseries{Construction}    & $\mathbf{N_{it}}$  \\
	\ref{lab:Ray_5G_5Iter}   &  5G Polar \cite{polar5G2018}      & 5         \\
	\ref{lab:Ray_NN_5Iter}   &  NN-designed    & 5         \\
	\ref{lab:Ray_5G_20Iter}  &  5G Polar \cite{polar5G2018}      & 20        \\	
	\ref{lab:Ray_NN_20Iter}  &  NN-designed    & 20        \\
	\ref{lab:Ray_5G_200Iter} &  5G Polar \cite{polar5G2018}      & 200       \\
};

\end{tikzpicture}%
} 
	\vspace{-0.6cm}
	
	\caption{\footnotesize BER performance of the \emph{learned} 
		$\mathcal{P}(256,128)$-code over the Rayleigh fading channel 
		under BP decoding with number of iterations $N_{it}$.}\label{fig:BER_ray_comb}	
\end{figure}

\section{Conclusion and Future avenues} \label{sec:conc}

A deep learning-based polar code design framework was introduced by inherently taking the actual decoder and channel into account. We have showcased that the \ac{DL} toolkit itself can be not just used to completely learn new structures but also to support \emph{classical} system design, namely the selection of frozen bit-positions of a polar code. This renders \ac{DL} techniques into a universal toolkit that deserves further analysis of its capabilities but also its limitations in the context of communications.

The major constraint of the binary vector representation of the information/frozen sets was relaxed to a soft-valued vector which can be regarded as training weights of a neural network. Proper loss functions and corresponding weights were also proposed to satisfy various design constraints (e.g., BER performance, code rate, etc.) while tailoring the code to the decoder and the channel. Results were shown for BP decoding under limited iteration count over both AWGN and Rayleigh fading channels with considerable performance gains over state-of-the-art construction schemes. Extending the current framework to setups with different parameters (e.g.,  codelengths, channels, decoders, etc.)  is straightforward.
The reason why we focused on the BP decoder-tailored design is (besides the easy access of the gradients) that, to the best of our knowledge, it still misses a systematic/analytical approach. 
It is also possible to deploy the proposed framework to design codes other than polar codes (e.g., LDPC codes) given the binary-to-soft conversion, rate adjustment, LLR saturation and other mentioned schemes. This is of particular significance for short codes, short LDPC codes for instance, that miss an analytical understanding and, thus, deep learning can prove to be most useful here.
Several future aspects might be interesting to look into, e.g., the impact of different rate adjustment schemes and the effect of optimizing the intermediate loss functions on the overall training performance.

\bibliographystyle{IEEEtran}
\bibliography{references}

\begin{thebibliography}{10}
\providecommand{\url}[1]{#1}
\csname url@samestyle\endcsname
\providecommand{\newblock}{\relax}
\providecommand{\bibinfo}[2]{#2}
\providecommand{\BIBentrySTDinterwordspacing}{\spaceskip=0pt\relax}
\providecommand{\BIBentryALTinterwordstretchfactor}{4}
\providecommand{\BIBentryALTinterwordspacing}{\spaceskip=\fontdimen2\font plus
\BIBentryALTinterwordstretchfactor\fontdimen3\font minus
  \fontdimen4\font\relax}
\providecommand{\BIBforeignlanguage}[2]{{%
\expandafter\ifx\csname l@#1\endcsname\relax
\typeout{** WARNING: IEEEtran.bst: No hyphenation pattern has been}%
\typeout{** loaded for the language `#1'. Using the pattern for}%
\typeout{** the default language instead.}%
\else
\language=\csname l@#1\endcsname
\fi
#2}}
\providecommand{\BIBdecl}{\relax}
\BIBdecl

\bibitem{nachmani2016learning}
E.~Nachmani, Y.~Be'ery, and D.~Burshtein, ``Learning to decode linear codes
  using deep learning,'' in \emph{54th Annual Allerton Conf. on Commun.,
  Control, and Computing}, 2016, pp. 341--346.

\bibitem{yihan2019deepturbo}
Y.~{Jiang}, H.~{Kim}, H.~{Asnani}, S.~{Kannan}, S.~{Oh}, and P.~{Viswanath},
  ``Deepturbo: Deep turbo decoder,'' \emph{arXiv:1903.02295}, 2019.

\bibitem{kim2018communication}
H.~Kim, Y.~Jiang, R.~Rana, S.~Kannan, S.~Oh, and P.~Viswanath, ``Communication
  algorithms via deep learning,'' \emph{arXiv:1805.09317}, 2018.

\bibitem{jiang2019mind}
Y.~Jiang, H.~Kim, H.~Asnani, and S.~Kannan, ``Mind: Model independent neural
  decoder,'' \emph{arXiv:1903.02268}, 2019.

\bibitem{gruber2017deep}
T.~Gruber, S.~Cammerer, J.~Hoydis, and S.~ten Brink, ``On deep learning-based
  channel decoding,'' in \emph{CISS}, 2017, pp. 1--6.

\bibitem{lyu2018performance}
W.~Lyu, Z.~Zhang, C.~Jiao, K.~Qin, and H.~Zhang, ``Performance evaluation of
  channel decoding with deep neural networks,'' in \emph{ICC}, 2018, pp. 1--6.

\bibitem{jiang2018learn}
Y.~Jiang, H.~Kim, H.~Asnani, S.~Kannan, S.~Oh, and P.~Viswanath, ``Learn codes:
  Inventing low-latency codes via recurrent neural networks,''
  \emph{arXiv:1811.12707}, 2018.

\bibitem{samuel2017deep}
N.~Samuel, T.~Diskin, and A.~Wiesel, ``Deep {M}{I}{M}{O} detection,'' in
  \emph{SPAWC}, 2017, pp. 1--5.

\bibitem{farsad2017detection}
N.~Farsad and A.~Goldsmith, ``Detection algorithms for communication systems
  using deep learning,'' \emph{arXiv:1705.08044}, 2017.

\bibitem{oshea2017introduction}
T.~O’Shea and J.~Hoydis, ``An introduction to deep learning for the physical
  layer,'' \emph{IEEE Trans. on Cognitive Commun. and Networking}, vol.~3,
  no.~4, pp. 563--575, 2017.

\bibitem{cammerer2017scaling}
S.~Cammerer, T.~Gruber, J.~Hoydis, and S.~ten Brink, ``Scaling deep
  learning-based decoding of polar codes via partitioning,'' in \emph{IEEE
  Global Commun. Conf. (GLOBECOM)}, 2017, pp. 1--6.

\bibitem{ArikanMain}
E.~{Arıkan}, ``{Channel Polarization: A Method for Constructing
  Capacity-Achieving Codes for Symmetric Binary-Input Memoryless Channels},''
  \emph{IEEE Trans. Inf. Theory}, vol.~55, no.~7, pp. 3051--3073, July 2009.

\bibitem{polar5G2018}
\BIBentryALTinterwordspacing
``{Technical Specification Group Radio Access Network},'' \emph{3GPP, 2018, TS
  38.212 V.15.1.1.} [Online]. Available:
  \url{http://www.3gpp.org/ftp/Specs/archive/38_series/38.212/}
\BIBentrySTDinterwordspacing

\bibitem{talvardyList}
I.~Tal and A.~Vardy, ``{List Decoding of Polar Codes},'' \emph{IEEE Trans. Inf.
  Theory}, vol.~61, no.~5, pp. 2213--2226, May 2015.

\bibitem{AdaptiveList}
B.~Li, H.~Shen, and D.~Tse, ``{An Adaptive Successive Cancellation List Decoder
  for Polar Codes with Cyclic Redundancy Check},'' \emph{IEEE Commun. Lett.},
  vol.~16, no.~12, pp. 2044--2047, Dec. 2012.

\bibitem{PSCL_Journal}
S.~A. Hashemi, M.~Mondelli, S.~H. Hassani, C.~Condo, R.~L. Urbanke, and W.~J.
  Gross, ``{Decoder Partitioning: Towards Practical List Decoding of Polar
  Codes},'' \emph{IEEE Trans. Commun.}, vol.~66, no.~9, pp. 3749--3759, Sep.
  2018.

\bibitem{ArikanBP}
E.~Arıkan, ``{A Performance Comparison of Polar Codes and Reed-Muller
  Codes},'' \emph{IEEE Commun. Lett.}, vol.~12, no.~6, pp. 447--449, June 2008.

\bibitem{elkelesh2018belief}
A.~Elkelesh, M.~Ebada, S.~Cammerer, and S.~ten Brink, ``{Belief Propagation
  List Decoding of Polar Codes},'' \emph{IEEE Commun. Lett.}, vol.~22, no.~8,
  pp. 1536--1539, Aug. 2018.

\bibitem{GenAlg_Journal_IEEE}
A.~{Elkelesh}, M.~{Ebada}, S.~{Cammerer}, and S.~{ten Brink},
  ``{Decoder-Tailored Polar Code Design Using the Genetic Algorithm},''
  \emph{IEEE Trans. Commun.}, vol.~67, no.~7, pp. 4521--4534, July 2019.

\bibitem{GenAlgLDPCDecInTheLoop}
A.~Elkelesh, M.~Ebada, S.~Cammerer, L.~Schmalen, and S.~ten Brink,
  ``{Decoder-in-the-Loop: Genetic Optimization-based LDPC Code Design},''
  \emph{IEEE Access}, 2019.

\bibitem{hershey2014deep}
J.~R. Hershey, J.~L. Roux, and F.~Weninger, ``Deep unfolding: Model-based
  inspiration of novel deep architectures,'' \emph{arXiv:1409.2574}, 2014.

\bibitem{Urbanke_chCsC_BP}
N.~Hussami, S.~B. Korada, and R.~Urbanke, ``{Performance of Polar Codes for
  Channel and Source Coding},'' in \emph{IEEE Inter. Symp. Inf. Theory (ISIT)},
  June 2009, pp. 1488--1492.

\bibitem{multi_trellis}
A.~Elkelesh, M.~Ebada, S.~Cammerer, and S.~ten Brink, ``{Belief Propagation
  Decoding of Polar Codes on Permuted Factor Graphs},'' in \emph{IEEE Wireless
  Commun. and Networking Conf. (WCNC)}, Apr. 2018.

\bibitem{sparseBP}
S.~{Cammerer}, M.~{Ebada}, A.~{Elkelesh}, and S.~{ten Brink}, ``{Sparse Graphs
  for Belief Propagation Decoding of Polar Codes},'' in \emph{IEEE Inter. Symp.
  Inf. Theory (ISIT)}, June 2018, pp. 1465--1469.

\bibitem{constructDE}
R.~Mori and T.~Tanaka, ``{Performance of Polar Codes with the Construction
  using Density Evolution},'' \emph{IEEE Commun. Lett.}, vol.~13, no.~7, pp.
  519--521, July 2009.

\bibitem{constructGaussian}
P.~Trifonov, ``{Efficient Design and Decoding of Polar Codes},'' \emph{IEEE
  Trans. Commun.}, vol.~60, no.~11, pp. 3221--3227, Nov. 2012.

\bibitem{constructTalVardy}
I.~Tal and A.~Vardy, ``{How to Construct Polar Codes},'' \emph{IEEE Trans. Inf.
  Theory}, vol.~59, no.~10, pp. 6562--6582, Oct. 2013.

\bibitem{BetaIngmard}
G.~He, J.~C. Belfiore, I.~Land, G.~Yang, X.~Liu, Y.~Chen, R.~Li, J.~Wang,
  Y.~Ge, R.~Zhang, and W.~Tong, ``{$\beta$-expansion: A Theoretical Framework
  for Fast and Recursive Construction of Polar Codes},'' in \emph{IEEE Global
  Commun. Conf. (GLOBECOM)}, Dec. 2017, pp. 1--6.

\bibitem{UPO}
C.~Schürch, ``{A Partial Order For the Synthesized Channels of a Polar
  Code},'' in \emph{IEEE Inter. Symp. Inf. Theory (ISIT)}, July 2016, pp.
  220--224.

\bibitem{BP_LLR_Siegel}
M.~Qin, J.~Guo, A.~Bhatia, A.~G. i~Fabregas, and P.~Siegel, ``{Polar Code
  Constructions Based on LLR Evolution},'' \emph{IEEE Commun. Lett.}, vol.~21,
  no.~6, pp. 1221--1224, June 2017.

\bibitem{RMurbankePolar}
M.~Mondelli, S.~H. Hassani, and R.~L. Urbanke, ``{From Polar to Reed-Muller
  Codes: A Technique to Improve the Finite-Length Performance},'' \emph{IEEE
  Trans. Commun.}, vol.~62, no.~9, pp. 3084--3091, Sep. 2014.

\bibitem{polarDesign5G}
V.~Bioglio, C.~Condo, and I.~Land, ``{Design of Polar Codes in 5G New Radio},''
  \emph{ArXiv e-prints}, Apr. 2018.

\bibitem{doan2018neural}
N.~Doan, S.~A. Hashemi, E.~N. Mambou, T.~Tonnellier, and W.~J. Gross, ``{Neural
  Belief Propagation Decoding of CRC-Polar Concatenated Codes},'' \emph{arXiv
  e-prints}, 2018.

\bibitem{hinton2012coursera}
G.~Hinton, ``Neural networks for machine learning,'' \emph{Coursera, video
  lectures}, 2012.

\bibitem{hubara2016binarized}
I.~Hubara, M.~Courbariaux, D.~Soudry, R.~El-Yaniv, and Y.~Bengio, ``Binarized
  neural networks: Training neural networks with weights and activations
  constrained to +1 or -1,'' \emph{arXiv:1602.02830}, 2016.

\bibitem{jaderberg2017population}
M.~Jaderberg \emph{et~al.}, ``Population based training of neural networks,''
  \emph{arXiv:1711.09846}, 2017.

\bibitem{hoeher2000log}
P.~Hoeher, I.~Land, and U.~Sorger, ``Log-likelihood values and {M}onte {C}arlo
  simulation-some fundamental results,'' in \emph{2nd Inter. Symp. on Turbo
  Codes and Related Topics}, 2000, pp. 43--46.

\bibitem{aoudia2018end}
F.~A. Aoudia and J.~Hoydis, ``End-to-end learning of communications systems
  without a channel model,'' in \emph{52nd Asilomar Conf. on Signals, Systems,
  and Computers}, 2018, pp. 298--303.

\bibitem{TrifonovRayleigh}
P.~Trifonov, ``{Design of Polar Codes for Rayleigh Fading Channel},'' in
  \emph{Inter. Symp. Wireless Commun. Syst. (ISWCS)}, Aug. 2015, pp. 331--335.

\end{thebibliography}

\end{NoHyper}
\end{document}